# Extremely large and significantly anisotropic magnetoresistance in ZrSiS single crystals

Yang-Yang Lv,[1†] Bin-Bin Zhang,[1†] Xiao Li,[2†] Shu-Hua Yao,[1*] Y. B. Chen,[2*] Jian Zhou,[1*] Shan-Tao Zhang,[1] Ming-Hui Lu,[1] Yan-Feng Chen[1,3]

[1]National Laboratory of Solid State Microstructures & Department of Materials Science and Engineering &, Nanjing University, Nanjing 210093 China

[2]National Laboratory of Solid State Microstructures & Department of Physics, Nanjing University, Nanjing 210093 China

[3]Collaborative Innovation Center of Advanced Microstructure, Nanjing University, Nanjing, 210093 China

[†]Yang-Yang Lv, Bin-Bin Zhang and Xiao Li contribute equally.

Corresponding authors: S. H. Y, shyao@nju.edu.cn; Y. B. C, ybchen@nju.edu.cn; J. Z, zhoujian@nju.edu.cn

**Abstract**

Recently, the extremely large magnetoresistance observed in transition metal telluride, like WTe$_2$, attracted much attention because of the potential applications in magnetic sensor. Here we report the observation of extremely large magnetoresistance as $3.0\times10^4$ % measured at 2 K and 9 T magnetic field aligned along [001]-ZrSiS. The significant magnetoresistance change ($\sim 1.4\times10^4$ %) can be obtained when the magnetic field is titled from [001] to [011]-ZrSiS. These abnormal magnetoresistance behaviors in ZrSiS can be understood by electron-hole compensation and the open orbital of Fermi surface. Because of these superior MR properties, ZrSiS may be used in the novel magnetic sensors.

Recently, materials scientists observed the extremely large magnetoresistance (MR) in a series of transition metal chalcogenides/telluride, rare-earth monopnictides, for example, WTe$_2$, LaSb.[1-3] These studies inspire the hot research on exploring the similar compounds with superior MR properties. From the view point of application, extremely large and sensitive MR are the basic requirements for magnetic memory/sensor devices.[4-6] As far as the basic science is concerned, what is the physical origin of extremely large MR in these materials? In addition, there are some strange fermions in these materials, such as Weyl fermion or Dirac nodal line at the reciprocal space.[7-12] A natural question is whether the extremely large MR is related to these un-conventional fermions? Bearing these considerations, we pay our attention to a new material-ZrSiS. In accordance to the theoretical prediction, ZrSiS is a semimetal, and maybe evolve to a weak/strong topological insulator under the external stimuli.[13] A recent angle resolved photoemission spectroscopy study claims that there is a Dirac nodal line feature in ZrSiS.[14] Then we asked several questions. Such as, can we observe the extremely large MR in ZrSiS too? Is MR behavior in ZrSiS different from that in WTe$_2$ that comes from the resonant electron-hole compensation,[1] or MR is only related to Fermi surface topology?[15] To answer these questions, we synthesized the ZrSiS single crystals and measured its magneto-transport property. As described as follows, we observe the unsaturated extremely large MR ($3.0\times10^4$ % at 2 K and 9 Tesla magnetic field aligned along [001]-ZrSiS) in ZrSiS, and significant MR change (~$1.4\times10^4$ % at 2 K under 9 Tesla magnetic field) when the magnetic field is tilted from [001] to [011]-ZrSiS. The MR is still kept as 11% at the ambient condition. The unsaturated, extremely large and significantly anisotropic MR in ZrSiS can be attributed to the electron-hole compensation, as well as open orbitals in Fermi surface of ZrSiS. Because of these superior MR properties, ZrSiS may be used in the novel Hall/MR magnetic sensors.

The single crystals ZrSiS were grown by a chemical vapor transport method using iodine ($I_2$) as a transport agent. The growth procedure includes two steps. Firstly, the polycrystalline ZrSiS powders were synthesized by the direct solid-state reaction using elemental Zr (GRINM 99.999%), Si (Alfa Aesar 99.999%) and S (Aladdin 99.99%) in sealed and evacuated (P~$4\times10^{-6}$ Torr) quartz ampoules. Prior to the reactions, the high purity quartz ampoules were immersed in a HCl + $HNO_3$ (3:1) solution for 3 h, then rinsed with de-ionized water and ethyl alcohol, finally dried at 673 K for 10 h. The stoichiometric amounts of ternary mixture were heated to 1373 K in steps of 100 K/h and maintained at this temperature for 7 days in a tube furnace till the reaction was complete. Then the samples were quenched to room temperature and the resulting single phase products were obtained. Secondly, the as-prepared polycrystalline ZrSiS powders together with $I_2$ in a concentration of 8 mg/cm$^3$ were loaded into the sealed evacuated quartz tube, and then placed at a two-zone furnace to synthesize crystals. After growing in a temperature profile of 1423-1273 K over a period of 10 days, the metallic luster crystals with the size up to $7\times2\times0.5$ mm$^3$ were successfully obtained. Single crystal X-ray diffraction (XRD) (Ultima III Rigaku X-ray diffractometer, Cu-$K\alpha$ radiation) with 2θ scanned from 5º to 75º was applied to investigate growth orientation of the polished crystals. The elemental compositions of the crystal samples were determined by an energy dispersive spectroscopy (EDS) spectrometer equipped in a scanning electron microscope (SEM) (FEI-Quanta) with the internal calibration. Standard six-probe method was employed on the roughly rectangular crystals for the electrical transport measurements which were performed in a 9 Tesla physical properties measurement system (PPMS, Quantum Design). During the electrical measurement, the electrical current is aligned along [010]-direction of ZrSiS.

The crystal of ZrSiS has PbFCl-type structure with tetragonal P4/nmm space group (No. 129).[16] As shown in Fig. 1(a), in the crystal structure of ZrSiS, each Zr atom is surrounded by four Si atoms and four S atoms from the square nets bounding the Zr layer. And the Si square nets and S square nets are located at opposite sides of Zr layers. All these layers are situated at the *ab*-plane and stacked along the crystallographic *c*-axis of the tetragonal unit cell forming a layer-like crystal structure. The typical photograph of the as-grown ZrSiS single crystals is depicted in Fig. 1 (b), displaying that the crystals show rectangular plate-like shape and metallic luster. The reflections in the XRD patterns of the maximum exposed surface of ZrSiS (see Fig. 1 (c)) can be indexed as (0 0 *k*)-ZrSiS. It suggests that the exposed surface of ZrSiS crystal belongs to *c*-plane. In addition, the full-width at half maximum of (001) pole of ZrSiS is as small as 0.07°, which substantiates the high crystalline quality of grown ZrSiS. The stoichiometry of ZrSiS is characterized by EDS. It substantiates that the ratio among Zr, Si and S is 1.00:1.02:0.96, which is quite close to stoichiometric one. The elemental uniformity is confirmed by EDS mappings that are presented in Fig. 1(d). As can be seen, the distribution of the Zr, Si and S elements in the sample is homogeneous. All the above-mentioned structural and chemical composition characterizations confirm that the as-grown ZrSiS crystal samples have single crystalline quality and are stoichiometric.

Fig. 2(a) shows the temperature-dependent resistivities in ZrSiS crystal under different magnetic fields that are aligned along [001]-ZrSiS. ZrSiS shows the metallic behavior from 2 K to 300 K without magnetic field. When the 9 Tesla magnetic field is applied along [001]-ZrSiS, one can see a metal-insulator transition around 110 K. And the resistivitives are saturated to a constant at the low temperature (~30 K). Fig. 2(b) shows the relationship between MR and magnetic field at several temperatures. There are three remarkable features in Fig. 2(b). Firstly, the MR of ZrSiS crystal is as large as 3.0×10⁴ % measured at 2 K and 9 Tesla magnetic field, which is comparable to that reported in WTe$_2$.[1] Secondly, the magnetic field dependent MR curves are parabolic-like at the whole measured magnetic ranges under these temperatures. Thirdly, there is Shubnikov-de-Hass oscillation at the low temperatures. It indicates the high mobility in ZrSiS.[1] Quantitatively, we used the two-carrier model to fit the magnetic-field dependent MR measured at 2 K. The MR in two-carrier model is[15]

$$MR = \frac{(|n_e|+|n_h|)^2 + (n_e+n_h)^2 \mu^2 B^2}{(n_e+n_h)^2 + (n_h-n_e)^2 \mu^2 B^2} - 1 \qquad (1)$$

where $n_e$, $n_h$, $\mu$ and $B$ are electron concentration, hole concentration, average mobility and magnetic field, respectively. We assume that mobility of electron is the same as that of hole. The fitting substantiates that ratio between electron and hole concentration is 0.91, as well as the average mobility is as large as 1.79×10⁴ cm²/(V·S). It suggests that ZrSiS also belongs to a semimetal similar to WTe$_2$.[1] Inset of Fig. 2(b) is the temperature-dependent MR under 9 T magnetic field. One can see that the MR of ZrSiS crystal still keeps quite large even at the high temperature. For example, the MR of ZrSiS crystals measured at 70 K and room temperature, under 9 Tesla magnetic field, still can reach 1450% and 11%, respectively. In comparison, the MR of WTe$_2$ at the same conditions only has 120% and <1.0%, respectively.[1] It may suggest that ZrSiS is more likely used in the real device applications compared to WTe$_2$.

To quantitatively discuss the metal-insulator transition in ZrSiS (see Fig. 2(a)), we defined two temperatures. One is the $T_m$ characterizing the temperature at which differentiation of resistivity over temperature changes from positive to negative, which is the indicator of metal-insulator transition; the other is temperature $T_I$ where the resistivity begins to saturate at low temperature (∼30 K). Fig. 3(a) shows the evolution of $T_m$ and $T_I$ as a function of magnetic fields for two samples. One can see that there is a magnetic field threshold to induce metal-insulator transition. The threshold magnetic field is around 2 Tesla. After it, $T_m$ is increased with increased magnetic field. Qualitatively, $T_m$ is increased from 30 K to 110 K when the magnetic field is increased from 0 T to 9 T. On the contrary, the $T_I$ is nearly independent on magnitude of magnetic field. Because $T_I$ is nearly insensitive to magnetic field magnitude, we think that magnetic field only induces the partial energy gap (or nominated as pseudo-gap[17]) on the electronic band structure of ZrSiS. Similar behavior is universally observed in unconventional superconductivity, or charge/spin-density wave insulators.[17]

To phenomenologically describe the metal-insulator transition induced by magnetic field, we used the Arrhenius equation ($\rho \propto \exp\left[-E_g/(k_B T)\right]$, where ρ, $E_g$, $k_B$ and $T$ are resistivity, energy gap, Boltzmann constant and absolute temperature, respectively.[18]) to fit the temperature-dependent-resistivity at temperatures ranged from $T_I$ to $T_m$. The relationship between extracted insulating gap and magnetic-field magnitude of two samples is plotted in Fig. 3(b). One can see that the extracted thermal active gaps at 9 T for two samples are 4.2 meV and 5.8 meV, respectively.

Fig. 4 shows the MR behavior of ZrSiS single crystal measured at 2 K and magnetic field scanned from -9 to 9 Tesla, and direction of magnetic field tilted from [001]- to [010]-ZrSiS. The definition of angle $\theta$ is indicated in the lower inset of Fig. 4(a). One can see two interesting features in Fig. 4(a). One is the parabolic relationship between MR and magnetic field in the whole measured magnetic-field range, the other is the significant MR anisotropy. The upper inset in Fig. 4(a) shows the angle-dependent MR in ZrSiS measured at 2 K and 9 T. One can see that the largest MR (~4.4×10$^4$ %) is observed when the magnetic field is aligned along [011]-direction of ZrSiS. When θ is smaller than 45º, the MR remains as large as 10$^4$ % order.

However, the MR decreases quickly when θ is larger than 45°. For example, the MR is only 1600 % at θ being 90°. In order to show anisotropic MR of ZrSiS at 2 K and 9 T more clearly, Fig. 4(b) demonstrates the polar diagram, which is a four-leaf-like pattern. It clearly shows that MR change can reach (∼$1.4\times10^4$ %) when the magnetic field is titled from [001] to [011]-ZrSiS. The significantly anisotropic MR is advantageous to magnetic sensor devices.[5,6] As to why the maximum MR observed in [011]-ZrSiS is under study now.

How can we understand the physical origin of the unsaturated, extremely large and significantly anisotropic MR observed in ZrSiS single crystals? Here we propose a possible explanation on this problem. Firstly, we assume that the Fermi surface in ZrSiS only has partially changes when the magnetic field is applied, the major topology feature of Fermi surface of ZrSiS still can be described by the single-electron approximation. The electronic band structure of ZrSiS, calculated by the density functional theory implemented in the VASP code,[19,20] is shown in Fig. 5(a). It can be seen that there are co-existence of electron and hole at the Fermi surface. The Fermi level of experimental ZrSiS crystal is determined by the comparison of Shubnikov-de-Hass oscillation frequency to theoretical Fermi surface extreme (not shown here). Then, we can draw the Fermi surface of ZrSiS (see Fig. 5(b)). As shown in Fig. 5(b), there are two features. Firstly, there are four branches of electron and holes at the Fermi surface, and hole channels show "cylinder-like" morphology. Quantitatively, the calculated ratio between electron and hole concentration is 0.95 that is quite close to experimental one (∼0.91, see discussion of Fig. 2). It suggests that ZrSiS is a semi-metal that is similar to $WTe_2$.[1] Secondly, when magnetic field is parallel to the [001]-direction of ZrSiS, all electron and hole orbitals under Lorentz force are closed ones; however, there are open electron and hole orbitals when the magnetic field is aligned along [010]-direction.[15] Based on the Fermi surface topology, it may suggest that the extremely large MR with magnetic field aligned along to [001]-ZrSiS is mainly attributed to resonant electron-hole compensation, which is quite similar to that in WTe$_2$.[1] On the contrary, when the magnetic field is applied along [011]-direction, the extremely large MR is attributed to both electron-hole compensation and open-orbital topology at Fermi surface. In addition, the non-saturated and parabolic MR behavior shown in Fig. 2(b) & 4(a) is easily understood. Because in either compensated semimetal or materials with Fermi surface topology containing open orbitals along certain directions, MR is not saturated and parabolically dependent on magnetic field.[21] Therefore, the extremely large, non-saturated and anisotropic MR observed in ZrSiS can be *qualitatively* understood by considering the Fermi surface topology of ZrSiS. It may be suggested that the extremely large MR of ZrSiS is not related to Dirac Fermion in ZrSiS.[14]

In the conclusion, we reported the observation of extremely large, un-saturated and significantly anisotropic MR in ZrSiS. The MR in ZrSiS can be reached as $3.0 \times 10^4$ % at 2 K and 9 Tesla magnetic field aligned along [001]-ZrSiS. The significant MR change ($\sim 1.4 \times 10^4$ %) can be observed when the magnetic field is tilted from [001] to [011]-ZrSiS. These "abnormal" MR behaviors of ZrSiS can be understood by considering the coexistence of electron and hole at Fermi surface, as well as the Fermi surface topology of ZrSiS. The MR of ZrSiS crystals measured at 70 K and room temperature, under 9 Tesla magnetic field, still can reach 1450% and 11%, respectively. It may be suggested that ZrSiS is more likely used in the novel Hall/MR magnetic sensors.

**Acknowledgements**

We would like to acknowledge the financial support from the National Natural Science Foundation of China (51472112, 51032003, 11374140, 11374149, 10974083, 11004094，11134006, 11474150 and 11174127), and major State Basic Research Development Program of China (973 Program) (2015CB921203, 2014CB921103, 2015CB659400). Y. -Y. Lv acknowledges the financial support from the Graduate Innovation Fund of Nanjing University (2015CL11).

**Figure captions**

**FIG. 1.** (a) The crystal structure of ZrSiS. (b) The typical photograph of the as-grown ZrSiS single crystals. (c)The XRD patterns of representative ZrSiS single crystal. (d) SEM image and Zr (green), Si (red) and S (blue) element mapping images of ZrSiS single crystal.

**FIG. 2.** (a) Temperature-dependent resistivities of ZrSiS crystal under different magnetic fields. When the magnetic field is applied, the metal-insulator transition happens at $T_m$ and the resistivities are saturated to a constant below $T_I$. Inset is the schematic of the experiment. (b) The relationship between magnetoresistance (MR) and magnetic field of ZrSiS single crystal measured at several temperatures, with magnetic field applied along the [001]-ZrSiS. Inset shows the temperature-dependent MR measured at 9 Tesla magnetic field aligned along [001]-ZrSiS.

**FIG. 3.** (a) The color picture for the resistivities are plotted as the function of temperatures and magnetic fields. And the temperature and magnetic field dependent $T_m$ and $T_I$ are also plotted. $T_m$ dramatically changes under magnetic field, while $T_I$ is nearly a constant. As the field is decreased from 9 T to 0 T, the two temperatures seem to merge at about 30 K and 1.8 T. (b) The dependence of insulating gap of two samples on the magnetic field.

**FIG. 4.** (a) The relationship between the MR of ZrSiS and magnetic field under different θ angles. θ changing from 0 to 90° corresponds to magnetic field tilted from [001] to [010]-ZrSiS. Upper inset shows the angle-dependent MR of ZrSiS single crystal measured at 2 K and 9 Tesla magnetic field. Lower inset is the schematic of the experiment. (b) The stereographic diagram of anisotropic MR of ZrSiS single crystal measured at 2 K and 9 Tesla magnetic field.

**FIG. 5.** (a) Electronic band structure of ZrSiS considering spin-orbit coupling. We assume that a small pseudo-gap is formed under the magnetic field, but the overall change to the band structure

remains negligible. (b) The corresponding Fermi surface of ZrSiS. Purple and yellow color represents electron and hole, respectively. a*, b* and c* are the unit vectors at the reciprocal space.

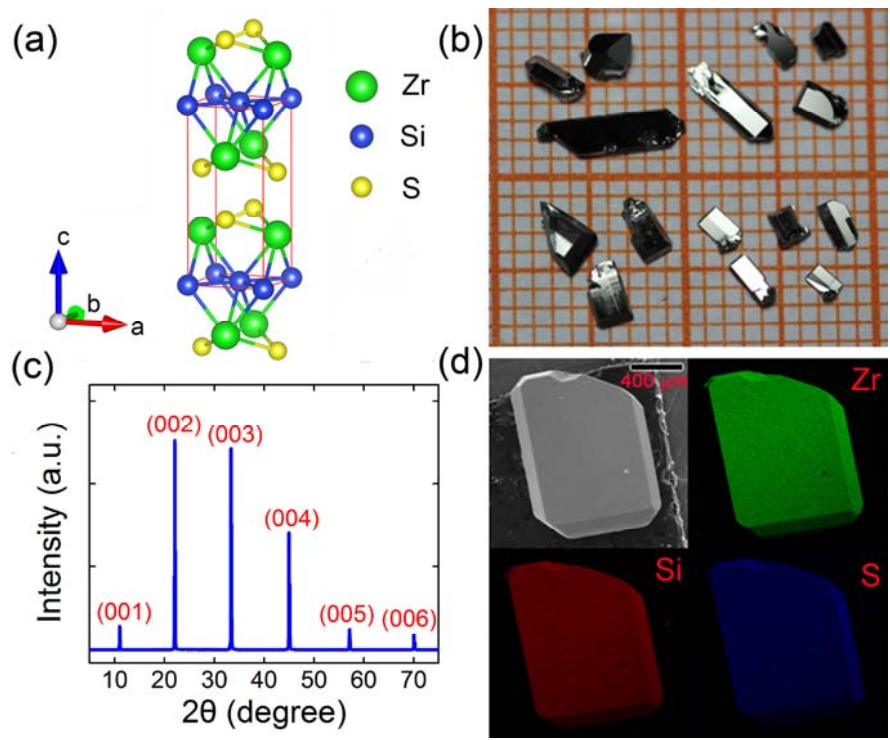

**FIG. 1**

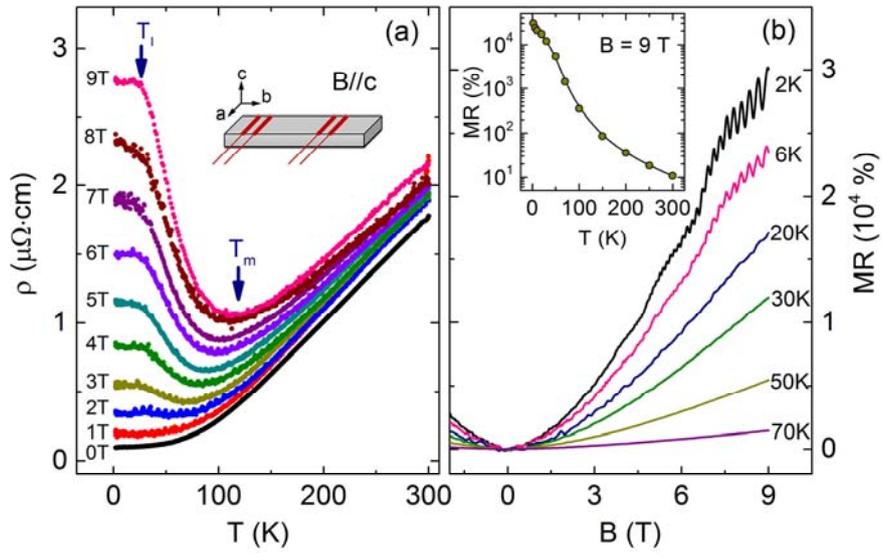

**FIG. 2**

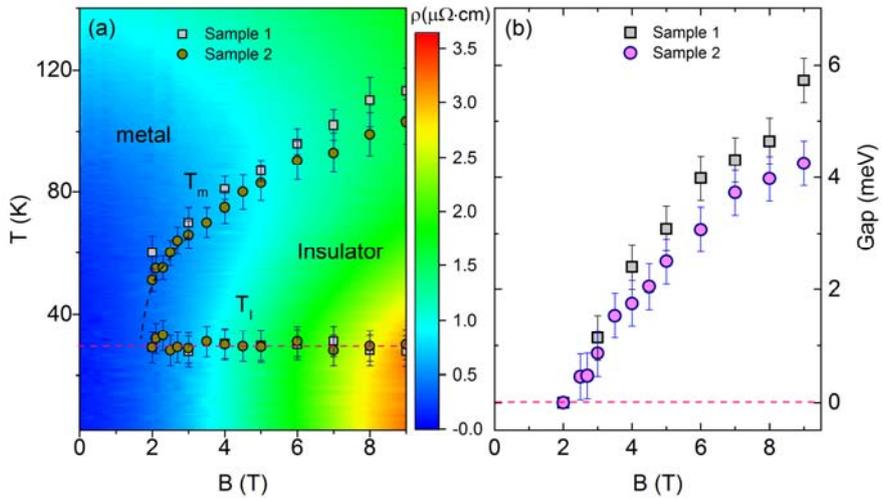

**FIG. 3**

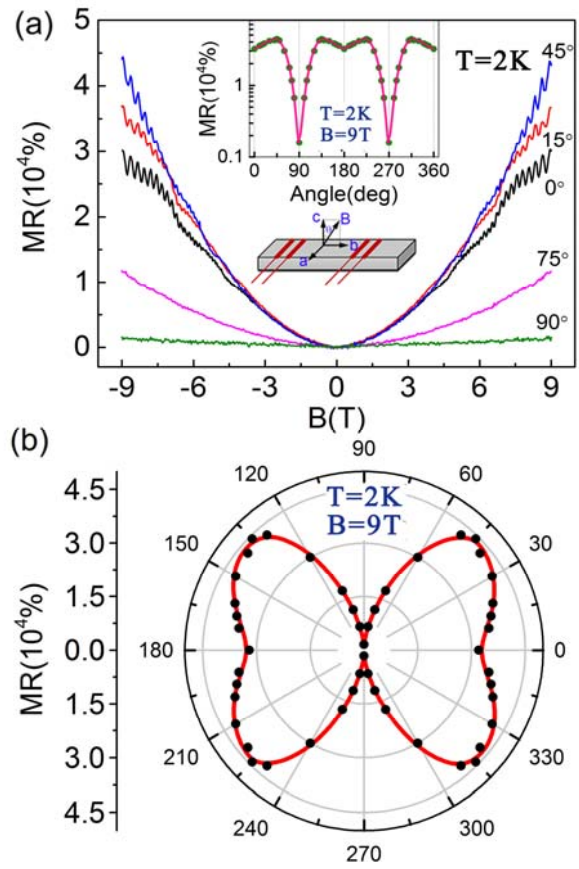

**FIG. 4**

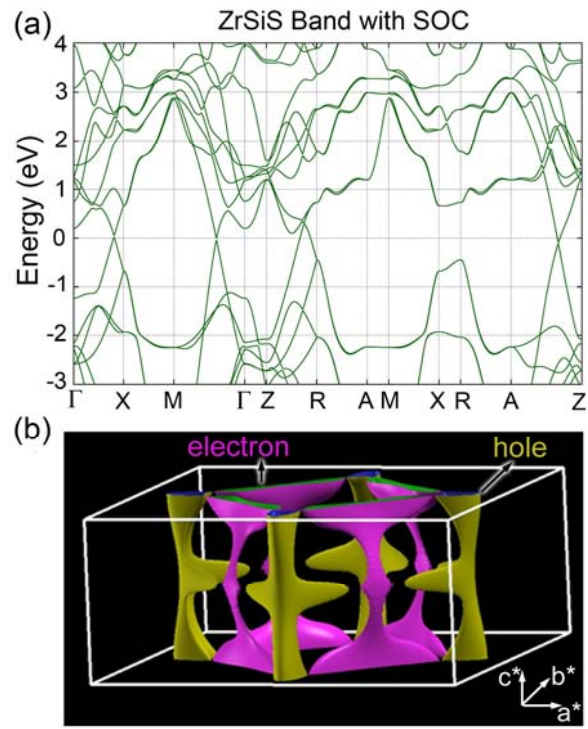

**FIG. 5**